\documentclass[showpacs,groupedaddress,
preprintnumbers,aps,prd]{revtex4}
\usepackage{amsmath,amssymb, amsthm,mathrsfs}
\def\be{\begin{equation}}
\def\ee{\end{equation}}
\def\bea{\begin{eqnarray}}
\def\eea{\end{eqnarray}}
\def\ba{\begin{array}}
\def\ea{\end{array}}
\newcommand{\bes}{\begin{subequations}}
\newcommand{\ees}{\end{subequations}}

\begin{document}
\preprint{}
\title{Kinetic surface roughening for the Mullins-Herring equation}

\author{Esmat Darvish$^{a}$}
\author{Amir Ali Masoudi$^{a,b}$}

\affiliation{$^{a}$Dep. of Physics, Islamic Azad University, Branch of North Tehran, P. O. Box 19585-936, Tehran, Iran \\
$^{b}$Department of Physics, Alzahra University, P. O. Box 19834, Tehran, Iran}


\begin{abstract}
Using the linearity property of the Mullins-Herring equation when the velocity is zero with
a Gaussian noise, we obtain an analytic form for the global mean-square surface width and height-height correlation function. This can be used to read the critical exponents in any dimension. In particular for $d=1$
we show that although the surface is super rough the system exhibits Family-Vicsek scaling behavior.  
\end{abstract}


\maketitle

\section{Introduction}
In nature one can find a huge number of systems that develop a rough interface in the process of growing. Many of them have been  understood by the use of some tools from fractal geometry \cite{bar}, such as scaling analysis as well as through modeling with stochastic partial differential equations (SPDEs) and discrete models. While these concepts do not constitute an exclusive theoretical framework for surfaces in the physical world, they can be applied to gain a deeper understanding of many processes, for example, in biology. Due to the many possible important applications in medicine, tumor growth constitutes one of the most interesting subjects of study to which scaling analysis can be applied. Actually very important research on tumor growth has been recently carried out. Strong experimental evidence has indicated that tumor growth belongs to the molecular beam epitaxy (MBE) universality class \cite {{bpf},{bas}}. 

The MBE dynamics is characterized by a number of features that include a linear growth rate, the constraint of growth activity to the outer border of tumor, and surface diffusion at the growing edge; all of them have been observed experimentally. Stochastic models for tumoral growth have recently been considered in \cite{{ce},{cas}}.         

The continuum equation which describes the MBE universality class, known as the Mullins-Herring equation \cite {mh}, 
has the following SPDE:
\be
\frac{\partial h}{\partial t}= -\kappa \nabla^4 h+ F_{0}+\eta({\bf x},t),
\label{2}
\ee 
where $h$ is the interface height, and $\kappa$ is the surface diffusion coefficient.
$F_{0}$  has dimension of velocity and may be interpreted as the product of the mean cell radius and the 
cell division rate. $\eta({\bf x},t)$ is a Gaussian noise with zero mean and  
\be
\left\langle \eta({\bf x},t) \eta({\bf x'},t')\right\rangle =2D \delta^d ({\bf x}-{\bf x'})\delta(t-t').
\label{1}
\ee
The critical exponents can be extracted from the equation (\ref {2}) simply by power counting. 
For $d=1$, if we ignore for the moment the velocity $F_{0}$, and perform the transformations $x\rightarrow bx$, $t\rightarrow b^z t$, and $h\rightarrow b^\alpha h$, we will get the scale invariant system only if 
 $z=4$ and $\alpha=3/2$, indicating that the system is super rough. Actually in this particular case the 
system is characterized by a set of critical exponents given by $\alpha=3/2$, $\alpha_{loc}=1$, $z=4$, $\beta=3/8$, and $\beta^*=1/8$ \cite{jmc}. Interestingly enough it was found that this system with the above critical exponents 
is compatible with the experimental data obtained from tumor growth measurement\cite{{bpf},{bas}}.

 One nice feature of the Mullins-Herring equation is its simplicity; it is linear and can be solved exactly by means of Fourier transformation.

We will be interested in the dynamics of a $d$ dimensional surface, which will be described by the stochastic variable $h(x,t)$ giving the height of the surface at time $t$ above substrate position $x$. Starting from a initially flat substrate of linear size $L$ at time $t=0$, the global mean square width of the surface,
\be
w^2(L,t)=\left\langle \frac{1}{N} \sum_{{\bf x}}\left(h({\bf x},t)-\frac{1}{N}\sum_{{\bf x'}}h({\bf x'},t)\right)^2\right\rangle
\label{hh1}
\ee
with $N$ being the total number of substrate sites, satisfies \cite{fv}
\be
w^2(L,t)\sim\;L^{2\alpha} F\left(\frac{t}{L^z}\right).
\label{3}
\ee
For $t\ll L^z$ the $L$ dependence drops out, and we get $w^2(L,t)\sim t^{2\beta }$. On the other hand 
for $t\gg L^z$ the scaling function becomes constant, and the Eq. (\ref {3}) 
yields $w^2\sim L^{2\alpha}$ \cite {kpz}. 

Another interesting quantity we will study is the height-height correlation function
\be
C({\bf r},t)\equiv\left\langle \left|h({\bf r},t)-h(0,t)\right|^2\right\rangle,
\label{rrr}
\ee
which  scales in the same way as the square of the local width, in the form  of
\be
C({\bf{r}},t)\sim r^{2\alpha} \;{\hat f}_{d}\left(\frac{\kappa t}{r^z}\right)
\ee
where $r=|\bf{r}|$ and the scaling function ${\hat f}_d (u)$ behaves as
\be
{\hat f}_d(u)\sim
\left\{
\ba{lll}
&const\;\;\;\;\;\;\;\;\;\;\;\;&{\rm if}\;\;\;\;\;u\gg1
\cr &&\cr
 &u^{2\beta}\;\;\;\;\;\;&{\rm if}\;\;\;\;u\ll1
 \ea
 \right.
 \label{Z}
 \ee
Studying the asymptotic behavior of the height-height correlation function 
provides an alternative method to determine the critical exponents.
 
The aim of this article is to calculate the mean square surface width and the scaling form for Mullins-Herring equation. We will also find an exact expression for the height-height correlation function $C({\bf {r}},t)$ with an emphasis on the asymptotic behavior of the  correlation function. Using this expression we find the scaling function ${\hat f}_{d}(\frac{\kappa t}{r^z})$ for different dimensions $d$.  
In particular for $d=1$ where the model could be used to describe the tumor growth we show that
the surface is super rough and exhibits the Family-Vicsek behavior.

\section{THE LINEAR LANGEVIN EQUATION }
Let us start from the linear Langevin equation which can be read from (\ref{2}) by setting $F_0=0$
\be
\frac{\partial h}{\partial t}=-\kappa \nabla^4 h+\eta({\bf x},t),
\label{zz}
\ee
Although the exponents 
\be
\alpha=\frac{(4-d)}{2}, \;\;\;\;\;\;\;\;\;z=4
\label{aa}
\ee
which characterizes the dynamics scaling of this equation are well-known, it is interesting to determine the value of various quantities in the scaling regime. These values are related to each other through the parameters $\kappa$ and $D$ of the continuum model.

Using the Fourier transformation one can find a solution for  Eq.(\ref {zz}) as follows
\be
h({\bf k},\omega)=\frac{1}{\kappa k^4-i\omega} \eta({\bf k},\omega),
\label{bb}
\ee
where $k=\left|{\bf k}\right|$ and
\bea
h({\bf k},\omega)&=&\int d^{d}{\rm x} \;\;dt\;\;  e^{-i({\bf k}.{\bf x}-\omega t)}h({\bf x},t),\cr
\eta({\bf k},\omega)&=&\int d^{d}{\rm x}\;\; dt\;\; e^{-i({\bf k}.{\bf x}-\omega t)}\eta({\bf x},t).
\label{dd}
\eea
We consider a growth process starting from a flat substrate in which one should set $\eta({\bf x},t)=0 $ for $t<0$. The noise correlator in Fourier space then takes the form  
\be
\left\langle \eta({\bf k},\omega) \eta({\bf k'},\omega')\right\rangle=2 D (2\pi)^{d}\delta^{d}({\bf k}+{\bf k'})\int_{0}^{\infty} d\tau e^{i(\omega+\omega')\tau} 
\label{ff}
\ee
By making use of the Eqs. (\ref{dd}) and (\ref{ff})  and after performing the $\omega$ integrals, we get the following expression for the amplitude  $h({\bf k},t)$
\be 
\left\langle h({\bf k},t) h({\bf k'},t')\right\rangle=\frac{D}{\kappa k^4}(1-e^{-2\kappa k^4 t})(2 \pi)^d\delta^d({\bf k}+{\bf k'})\;\;\;\;\;\;\;\;\;for \;\;t\geq0
\label{gg}
\ee

\subsection{Mean square width}
In this subsection we obtain a closed form for the mean square width  which describes the interface fluctuations and study its
asymptotic behaviors. This could be used to read the critical exponents.

To proceed we note that the mean-square width (\ref{hh1}) can be recast to the following form
\be
w^2(L,t)=
\int \frac {d^{d}{\bf k}}{(2\pi)^d}\frac{d^d{{\bf k'}}}{(2\pi)^d}\left\langle h({\bf k},t) h({\bf k'},t)\right\rangle.
\label{hh}
\ee
 Performing the momentum integration over a spherical shell from  $\frac{2\pi}{L}$  to  $\frac{\pi}{a}$ and carrying out the angular integrations,  we arrive at
\be
w^2(L,t)=K_{d}\;\;\int_{\frac {2\pi }{L}}^\frac{\pi}{a}\;\;\frac {dk}{k^{5-d}}\;\frac{D}{\kappa}
(\;1-e^{-2\kappa k^4 t}),
\label{ii}
\ee 
where ${K_d}^{-1}\;=2^{d-1}\;\;\pi^{\frac{d}{2}}\Gamma(\frac{d}{2})$.  In comparison with the lattice model, $a$ should
be identified with the lattice spacing. Using integrating by parts, from Eq.(\ref{ii}) one finds
\be
w^2 (L,t)=\;A\;+\frac{D}{\kappa}\;L^{4-d}\;\;F_d(\frac {\kappa t}{L^4}) +{\cal O}(e^{-2 {\pi^4}\frac{\kappa t}{a^4}}),
\label{jj}
\ee
where
\be
A=\frac{K_d}{d-4}\left(\frac{D}{\kappa}\right){\left(\frac{\pi}{a}\right)}^{d-4}
\label{kk}
\ee
and
\be
F_d(x)=\frac{K_d}{(4-d)(2\pi)^{4-d}}\left(1-\;e^{(-32 {\pi}^4 x)}\;+\;(32 {\pi}^4 \;x)^{1-\frac{d}{4}} \int_{32{\pi}^4 x}^{\infty}\;y^{\frac{d}{4}-1}\;e^{-y}\; \;dy \right)
\label{ll}
\ee
For large $x$, $ F_d(x)\;$ approaches a constant, while at small $x$, the limiting form of Eq. (\ref{ll})  is given by,
\be
F_d(x)=\frac{\Gamma {(\frac{d}{4})}}{\Gamma{(\frac{d}{2})}}\;\;\frac{1}{4-d}\;2^{\frac{(8-5d)}{4}}\;\;\pi^{-\frac{d}{4}}\;\;x^{\frac{(4-d)}{4}}\;+\;{\cal O}(x)\;\;\;\;\;\;\;{\rm for}\;\;d>0
\label{mm}
\ee
leading to the exponents as in Eq. (\ref{aa}).

We note, however, that for $d=4$ special care is needed. Actually since in this case $\alpha$ and $\beta$ are zero
one needs to consider this case separately and indeed redo the computations from first step with
$d=4$. It is then straightforward to go through the computations to find 
\be
w^2(L,t)=\frac{D}{8 \pi^2\kappa}\;\ln \left[\frac{L}{a}\;{\hat F}\left(\frac{\kappa t}{L^4}\right)\right] \;+\;{\cal O}\left(e^{\frac{-2 \pi^4\kappa t}{a^4}}\right)
\label{nn}
\ee
where
\be
\ln \;{\hat F}(x)=\;-\;\ln2\;+\;\frac{1}{4} \;Ei\left({-32\;\pi^4\;x}\right)
\label{oo}
\ee
Here  $Ei(-y)=-\int_{y}^{\infty}\;u^{-1}\;e^{-u}\;du\;$  is the exponential integral. By making use of the asymptotic behavior of $ Ei(-y)$  for small $y$, one gets
\be
w^2(L,t)\sim
\left\{
\ba{ll}
 \frac{D}{32\pi^2\kappa}\left[\gamma+\ln\left(\frac{2\pi^4\kappa t}{a^4}\right)+{\cal O}\left(\frac{\kappa t}{L^4}\right)\right]\;\;\;\;\;\;\;\;\;&{\rm for}\;\;a\ll{(\kappa t)}^\frac{1}{4}\ll L  \cr
& \cr \frac{D}{8\pi^2 \kappa}\ln{(\frac{L}{a})}+{\cal O}(1)\;\;\;\;\;\;\;\;\;\;\;\;\;\;\;\;\;\;\;\;\;\;\;\;\;\;\;\;\;\;\;\;\;\;\;\;\;\;&{\rm for}\;\;\;\;{(\kappa t)}^\frac{1}{4}\gg L
\ea
\right.
\ee
In this case at early times the width scales logarithmically with time and the saturation width depends on the logarithm of the system size.

\subsection{Height-height correlation function}

The aim of this subsection is to reproduce the critical exponents using the height-height
correlation function. To do this we will first obtain a closed form for the height-height
correlation function, then we will read the form of the scaling function.
By making use of the asymptotic behavior of the scaling function we can read
the critical exponents. 
 
To proceed we note that by going to the momentum space and using Eq.  (\ref{gg}) 
the height-height correlation (\ref{rrr}) reads
\be
C({\bf r},t)=\int_{2\pi /L}^{\pi /a}\;\;\frac{d^{d} {\rm k}}{(2\pi )^d}\; \frac{2D}{\kappa k^4}\;\left(1-e^{-2\kappa k^4 t}\right)\times\bigg{[}1-\cos {({\bf k}.{\bf r})}\bigg{]}.
\ee
Carrying out the angular integration, one finds 
\be
C({\bf r},t)=K_d\int_{2\pi/L}^{\pi /a}\;\;dk\; k^{d-5}\;\frac{2D}{\kappa}\;\left(1-e^{-2\kappa  k^4 t}\right)\times\left[1-\Gamma\left(\frac{d}{2}\right)J_{d/2-1}\;\;\left(kr\right)\left(\frac{kr}{2}\right)^{1-\frac{d}{2}}\right],
\label{a}
\ee
where $r=\left|{\bf r}\right|$, and $J_{n}(x)$ is the $n$th-order Bessel function. In general we need
to evaluate this integral and study its asymptotic behaviors as a function of $r$. To do this we note that 
the integral can be broken into two parts. The first part is independent of $r$, and the $r$ dependence comes 
from the second part. More explicitly, assuming $r\ll L$ and $(\kappa t)^{1/4}\ll L$, one has
\be
C({\bf r},t)=K_d\;\frac{2D}{\kappa}\left[(\kappa t)^{\frac{4-d}{4}}
\int^{ {\pi(\kappa t)^{1/4}}/{a}}_{0} dz\; z^{d-5}\left(1-e^{-2z^4}\right) 
-r^{4-d}\int^{\pi r/a}_{0}\;dy\; y^{d-5}\frac{\Gamma(d/2)J_{d/2-1}(y)}{(2/y)^{1-d/2}}\left(1-e^{-2\frac{\kappa t}{r^{4}}y^{4}}\right)\right].
\label{cc}
\ee
Here we have used the change of variables $z=(\kappa t)^{1/4} k$ and $y=rk$. The first integral can be
carried out leading to 
\be
C({\bf r},t)=K_d\;\frac{2D}{\kappa}\left[\frac{a^{4-d}}{(d-4) \pi^{4-d}}-\frac{(\kappa t)^{\frac{4-d}{4}}}{2^{1+d/4}}
\bigg{(}\Gamma(d/4-1)-\Gamma(d/4-1,2\pi^4\kappa t/a^4)\bigg{)}\right]-K_d \frac{2D}{\kappa}\;r^{4-d}
{\cal G}_d\left(\frac{r}{a},\frac{\kappa t}{r^4}\right),
\ee
for $d\neq 4$. Here ${\cal G}_d$ is defined by the second integral in (\ref{cc}). For $a\ll r$ this function is independent 
of $a/r$ and the height-height correlation function can be recast to the following form
\be
C({\bf r},t)=\frac{2D}{\kappa} r^{4-d} \hat{f}_d\left(\frac{\kappa t}{r^4}\right)+{\cal O}(1),
\ee
where $\hat{f}_d(\kappa t/r^4)$ is the scaling function.

It is then easy to evaluate the scaling behavior of the height-height correlation function which 
for $d\neq 2$ is given by  
 \be
 C({\bf r},t)\sim
 \left\{
 \ba{lll}
 & K_d\;\frac{2D}{\kappa}\left[\frac{a^{4-d}}{(d-4)\pi^{4-d}} -2^{d-5}\Gamma(d/2)\Gamma(d/2-2) \;r^{4-d}\right]\;\;\;\;\;\;\;\;\;\;\;\;\;\;\;\;\;\;\;\;\;\;\;&{\rm for}\;\;\;a\ll r \ll{(\kappa t)}^{\frac{1}{4}} \cr &&\cr
 &K_d\;\frac{2D}{\kappa}\left[\frac{a^{4-d}}{(d-4)\pi^{4-d}}-\frac{\Gamma(d/4-1)}{2^{1+d/4}}
 (\kappa t)^{\frac{4-d}{4}}\right] \;\;\;\;\;\;\;\;\;\;\;\;&{\rm for}\;\;\;a\ll {(\kappa t)}^{\frac{1}{4}}\ll r
 \ea
 \right.
 \label{d}
 \ee
For $d=2$ the calculation yields 
\be
C({\bf r},t)\sim
\left\{
\ba{lll}
&\frac{D}{\kappa}\frac{r^2}{16\pi}\left({\cal O}(1)+\log\left(\frac{\kappa t}{r^4}\right)\right)\;\;\;\; &{\rm for}\;\;\;\;a\ll r\ll (\kappa t)^{\frac{1}{4}} \cr&& \cr
&\;\frac{D}{\pi \kappa}\left(\kappa t\right)^{1/2}\left({\cal O}(1)-{\cal O}\left(\frac{exp \left(\frac{3}{8}\left(\frac{\kappa t}{r^4}\right)^{-1/3}\right)}{\left(\frac{\kappa t}{r^4}\right)^{11/6}}\right)\right)\;\;\;\;\;\;\;&{\rm for}\;\;\;\;a\ll (\kappa t)^{\frac{1}{4}} \ll r
\ea
\right.
\label{ddd}
\ee
We note that there are two extra terms appearing in the asymptotic behaviors (logarithmic term for $\frac{\kappa t}{r^4}\gg 1$ and exponential growth for $\frac{\kappa t}{r^4}\ll 1$) which  make it impossible to find the scaling form  for this case.

We note also that in the case of $d=5$ the roughness exponent is negative indicating that
the interface is flat. Taking into account that for the case of $d=3$ the exponent is
positive one may conclude that there would be a phase transition for $d=4$ case. Indeed this
is exactly what we have observed from mean square width's computations. 
In fact for $d=4$ the Eq. (\ref {a}) reads
 \be
 C({\bf r},t)\;=\;\frac{1}{4\;\pi^2}\;\frac{D}{\kappa}\;\int_{0}^{\pi/a}\;dk \;k^{-1}\;\left(\;1-\;e^{-2 \kappa k^4\;t}\right)\left(\frac{2 J_{1}(kr)}{kr}\right)
 \label{i}
 \ee
This integral can be evaluated leading to the following limiting behavior 
 \be
 C({\bf r} ,t)\;=
 \left\{
 \ba{lll}
 &\frac{1}{4 \pi^2}\;\frac{D}{\kappa}\;\log{\frac{r}{a}}\;+\;{\cal O}(1)\;\;\;\;\;\;\;\;\;&{\rm for}\;\;\;\;\;a\ll r\ll (\kappa t)^{\frac{1}{4}}\cr &&\cr
 &\frac{1}{16 \pi^2 }\;\frac{D}{\kappa}\;\log(\frac{\kappa t}{a^4})\;+\;{\cal O}(1)\;\;\;\;\;\;\;\;&{\rm for}\;\;\;\;a\ll (\kappa t)^{\frac{1}{4}}\ll r
 \ea
 \right.
 \label{j}
 \ee
As we see the correlation function decays logarithmically and therefore we get the logarithmic behavior for the scaling function, as expected.

\section{Conclusion}
In this paper we have obtained explicit expressions for the mean square width  of the $d$ dimensional surface for the Mullins-Herring equation which in asymptotic behavior has the standard Family-Vicsek scaling form.
We have also considered the height-height correlation function and studied its asymptotic behavior
to read the critical exponents.

Although we have worked out the height-height correlation function for any dimension, the most interesting case 
is $d=1$ where from (\ref{d}) we find $\alpha=3/2,z=4$ and $\beta=3/8$. Actually it was
shown in \cite{{bpf},{bas}} that these exponents are compatible with the experimental data obtained from tumor growth measurement.
In other words one would expect that the one dimensional Mullin-Herreing equation can describe
the tumor growth. Therefore it can be utilized to study the roughness of the tumor surface 
using, for example, the height-height correlation function. In particular we have seen that since the global roughness exponent is bigger than one the surface is super rough and the height-height correlation saturates. This shows that the 
surface of tumor is super rough and exhibits the Family-Vicsek behavior. This has to be compared with the
numerical results presented in \cite{jmc} it was shown that for large $L$ the system does not display any anomaly 
indicating that scaling function has the Family-vicsek shape.

Finally, for completeness, let us also consider the auto height correlation in the limit  $t$, $t'\rightarrow\infty$,
 \be
 \left\langle \left|h({\bf r},t)\;-h({\bf r},t')\right|^2\right\rangle\ \;=\;K_{d}\int_{0}^{\pi/a}\;dk\;k^{d-5}\;\frac{2D}{\kappa}\times\;\left(1-e^{-\kappa k^4\;\left|t-t'\right|}\right)
 \label{k}
 \ee
 Performing the steps similar to the previous case we get
 \be
 \left\langle \left|h({\bf r},t)\;-\;h({\bf r},t')\right|^2\right\rangle\sim
 \left\{
 \ba{lll} &2A\;+\;\frac{D}{\kappa}\frac{2^{2-d}}{4-d}\pi^{-d/2}\;\frac{\Gamma(d/4)}{\Gamma(d/2)}\left(\kappa\;\left|t-t'\right|\right)^{(4-d)/4}\;\;\;\;\;\;\;\;&{\rm for}\;\;\;\;d\neq4\cr  &&\cr
 &\frac{D}{16\;\pi^2\;\kappa}\log\left(\frac{\pi^4\;\kappa\left|t-t'\right|}{a^4}\right)\;+\;{\cal O}(1)\;\;\;\;&{\rm for}\;\;\;\;d=4
 \ea
 \right.
 \label{l}
 \ee
where $A$ is given by Eq. (\ref {kk}). \\

{\bf {Acknowledgments}}\\We would like to thank Prof. M. R. Rahimi Tabar and Prof. M. Alishahiha for useful discussions.


\begin{thebibliography}{99}


\bibitem{bar}
A.-L.Barabasi and H. E. Stanley, Fractal concepts in surface Growth(Cambridge University press,Cambridge,1995). 
\bibitem{bpf}
A.Br\'u, J. M. Pastor, I. Fernaud, I. Br\'u, S. Melle, and C. Berenguer, Phys. Rev. Lett. ${\bf 81}$, 4008(1998)
\bibitem{bas}
A. Br\'u, S. Albertos, J. L. Subiza, J. L. Garc\'ia-Asenjo, and I. Br\'u, Biophys. J. {\bf 85}, 2948(2003).
\bibitem{ce}
C. Escudero, Phys. Rev. E {\bf 73}, 020902(R)(2006)
\bibitem{cas}
C. Escudero, Phys. Rev. E {\bf 74}, 021901(2006)
\bibitem{mh}
W. W. Mullins, J. Appl. Phys. {\bf 28}, 333 (1957); C. Herring, J. Appl. Phys. {\bf 21}, 301(1950)
\bibitem{jmc}
J. M. L\'opez, M. A. Rodr\'iguez, and R. Cureno, Physica A {\bf 246}, 329(1997)
\bibitem{fv}
F. Family and T. Vicsek, J. Phys. A {\bf 18},L75 (1985); R. Jullien and R. Botel, ibid. {\bf 18}, 2279(1985).
\bibitem{kpz}
T. Natterman, L.-H. Tang, Phys. Rev. A {\bf 45}, 7156(1991) 
\end{thebibliography}
\end{document}